\documentclass[twoside]{article}
\usepackage{fleqn,epsf}
\makeatletter

\newlength\aboveauthorskip
\newlength\aboveaddressskip
\def\@listI{\leftmargin\leftmargini
            \parsep 2\p@ plus1\p@ minus0.5\p@
            \topsep 4\p@ plus1\p@ minus2\p@
            \itemsep 2\p@ plus1\p@ minus0.5\p@}

\def\@listii {\leftmargin\leftmarginii
              \labelwidth\leftmarginii
              \advance\labelwidth-\labelsep
              \topsep    3\p@ plus1\p@ minus0.5\p@
              \parsep    1\p@ plus0.5\p@  minus0.5\p@
              \itemsep   \parsep}

\def\thetable{\Roman{table}}

\def\pacs#1{\def\@pacs{#1}}

\def\@authoraddress{}

\def\author#1{\expandafter\def\expandafter\@authoraddress\expandafter
{\@authoraddress %
{%
{
\nointerlineskip
\vskip\aboveauthorskip
\normalsize\sc
\ignorespaces#1
\par
}}}}

\def\address#1{\expandafter\def\expandafter\@authoraddress\expandafter
{\@authoraddress{
\nointerlineskip
\vskip\aboveaddressskip
\small\ignorespaces#1\par
}}}

\def\ps@firsthdr{
  \footskip=9.5mm
  \def\@oddhead{\noindent Vol.~{\bf 97}~(2000)\hfill {\sl ACTA PHYSICA
      POLONICA A}\hfill No. 1(2)}
  \def\@oddfoot{\centerline{ (\thepage )}}
  }

\def\ps@standardheadings{
  \def\@oddhead{\hfill\thepage}
  \def\@evenhead{\thepage\hfill}
  \def\@oddfoot{}
  \def\@evenfoot{}
  }

\renewcommand\maketitle{
  \thispagestyle{firsthdr}
  {\par  \parindent=-1.8cm  \rule{162mm}{0.32mm}\par}
  \vskip 6mm
  {\hfill \footnotesize Proceedings of the European Conference
                      ``Physics of Magnetism 99", Pozna\'n 1999}
  \vskip 7.5mm
  {\centering
    \baselineskip=7mm
    {\Large \@title }\par
    \vskip 0.5mm 
    \@authoraddress \par
    {\it (Version date: \@date)} \par
    \vskip 2mm \normalsize
    \setcounter{footnote}{0}}
  }

\renewenvironment{abstract} {%
  \begin{center}
    \begin{minipage}{107mm}
      \small
      \parindent 7mm
      }{\vskip 1mm PACS numbers: \@pacs \end{minipage}\end{center}\vskip 1mm}

\def\@seccntformat#1{\csname the#1\endcsname.~}
\renewcommand\thesection {\@arabic\c@section}
\renewcommand\thesubsection   {\thesection.\@arabic\c@subsection}
\renewcommand\thesubsubsection{\thesubsection.\@arabic\c@subsubsection}
\renewcommand\section{\@startsection {section}{1}{\z@}%
                                   {2mm plus 1mm minus0.5mm}%
                                   {2mm plus 1mm minus0.5mm}%
                                   {\centering \rm \bf}}
\renewcommand\subsection{\@startsection{subsection}{2}{\z@}%
                                      {2mm plus 1mm minus0.5mm}%
                                      {2mm plus 1mm minus0.5mm}%
                                     {\centering \it}}
\renewcommand\subsubsection{\@startsection{subsubsection}{3}{\z@}%
                                      {1mm plus 0.5mm minus0.2mm}%
                                      {1mm plus 0.5mm minus0.2mm}%
                                     {\it}}

\renewcommand\figurename{Fig.}
\renewcommand\tablename{TABLE}

\def\fnum@figure{\figurename~\thefigure}
\def\fnum@table{\tablename~\thetable}

\long\def\@makecaption#1#2{%
  \vskip 10pt
  \sbox\@tempboxa{\small #1. #2}%
  \ifdim \wd\@tempboxa >\hsize
    {\small #1. #2}\par
  \else
    \global \@minipagefalse
    \hb@xt@\hsize{\hfil\box\@tempboxa\hfil}%
  \fi
   \vskip 0pt}

\def\tabcaption#1#2{%
  \refstepcounter\@captype
  \parbox{#1}{\small \baselineskip=1.1\baselineskip \hfill
    TABLE~\thetable \\\noindent #2}
  \vskip 0.1cm}

\def\thebibliography#1{\section*{\refname}%
  \small%
  \list{\@biblabel{\@arabic\c@enumiv}}%
  {%
     \leftmargin\labelwidth
     \advance\leftmargin\labelsep
     \usecounter{enumiv}%
     \let\p@enumiv\@empty
     \renewcommand\theenumiv{\@arabic\c@enumiv}}%
  \sloppy
  \clubpenalty4000
  \@clubpenalty \clubpenalty
  \widowpenalty4000%
  \sfcode`\.\@m
  \itemsep=0.0mm
 }

\def\table{\renewcommand{\arraystretch}{1.1}
     \doublerulesep=1mm
     \intextsep=3mm
     \topsep=0mm
     \@float{table}}

\makeatother

\begin{document}            

\title{EQUATIONS OF MOTION FOR A CHARGED PARTICLE IN n-DIMENSIONAL
MAGNETIC FIELD} 

\author{Wojciech Florek and Marek Thomas}

\address{Institute of Physics, Adam Mickiewicz University
  \\  Umultowska 85, 61-614 Pozna\'n, Poland}

\date{April 28, 1999}

\maketitle                   


\begin{abstract}
 The equation of motion for a charged particle moving in the $n$-dimensional constant
magnetic filed is obtained for any linear gauge and any metric tensor by generalization
of Johnson and Lipmann's approach. It allows to consider the magnetic orbits in the
$n$-dimensional space. It is shown that the movement of a particle can always be
decomposed into a number of two-dimensional cyclotronic motions and a free particle part.
\end{abstract}

\section{Introduction} 

The motion of a charged particle in a constant magnetic field has
been considered by many authors since pioneering works of Landau and Peierls \cite{LP}.
Johnson and Lipmann \cite{JL} in their wide-cited paper treated this problem in both
relativistic and non-relativistic quantum approach, introducing in
fact operators for coordinates of the orbit center. Another important
series of papers was started by independent, but very closely
related, works of Brown \cite{EB} and Zak \cite{Zab} concerning a
Bloch electron in a magnetic field. 

This paper is inspired by a comment of Menon and Agrawal \cite{MA} in
which the $n$-dimensional magnetic field was determined.
We consider a general form of the $n$-dimensional vector potential {\bf A}
determining a constant and uniform magnetic field {\sf H}.
The equations of motion, analogous to those given by Johnson and Lipmann \cite{JL}, 
are obtained and they lead to decomposition of the coordinates into two parts
corresponding to
the cyclotron and free movements, respectively. In the first part,
the movement of the orbit center and the relative coordinates can  be
distinguished. 

\section{Linear gauge and equations of motion in general form}

The $n$-dimensional magnetic field can be defined as the antisymmetrization
of $\nabla{\bf A}$, with $\nabla$ denoting the $n$-dimensional gradient 
operator, i.e.\ \cite{MA}
 \begin{equation}\label{mag}
 {\sf H}_{jk}=\partial_j A_k-\partial_k A_j\,,
 \end{equation}
 where ${\bf A}=(A_1,A_2,\dots,A_n)$ is an $n$-dimensional
covariant function of the coordinates, ${\bf A}\equiv {\bf A}({\bf x})$,
which can be referred to as the $n$-dimensional vector potential. 
If ${\bf A}$ is a {\it linear} function of the
coordinates then it can be expressed by an $(n\times n)$-matrix {\sf A} 
(a covariant tensor) as 
 \begin{equation}\label{vAmA}
 A_j({\bf x})={\sf A}_{kj}x^k
 \end{equation}
 (the Einstein convention is assumed throughout). Substituting it to 
(\ref{mag}) one obtains that {\sf H} is simply the antisymmetrized tensor
{\sf A}, i.e. 
 \begin{equation}\label{maga}
  {\sf H}={\sf A}-{\sf A}^T\,,
 \end{equation}
 so {\sf H} corresponds to a constant and uniform magnetic field.
 Working in the radiation gauge the subsidiary condition
$\nabla\cdot{\bf A}=g^{jk}\partial_j A_k=g^{jk}{\sf A}_{jk}=0$ has to be 
assumed.  
In particular, this form includes the Landau gauge
and the antisymmetric
gauge determined by ${\sf A}={\sf H}/2$. 
By transposition of the matrix {\sf A} one obtains a magnetic field
${\sf H}^T$ associated with {\sf H}:
 \begin{equation}\label{magt}
 {\sf H}^T_{jk}={\sf A}^T_{jk}-{\sf A}^T_{kj}=-{\sf H}_{jk}\,.
 \end{equation}

The Hamiltonian for a free particle in an external magnetic field,
with the effective mass $m$ and the charge $q$, can be written as
  $$
 {\cal H}= \frac{1}{2m}{\bf p}^2 =\frac{1}{2m}g^{jk}p_jp_k \,,
 $$
 with ${\bf p}$ being the canonical momentum, i.e.\ an
$n$-dimensional covariant operator 
 \begin{equation}\label{pcan}
 p_j=-{\rm i}\hbar\partial_j-\frac{q}{c}{\sf A}_{kj}x^k\,.
 \end{equation}
 These operators satisfy the commutation relation
 \begin{equation}\label{comm}
 [p_j,p_k] = {\rm i}\hbar\frac{q}{c}{\sf H}_{jk}\,,
 \end{equation}
 so they do not commute for ${\sf H}_{jk}\neq0$. However, it is important in
the further considerations that these operators are numbers, i.e.\ commute
with any operator. Equations of motion
for $p_j$ and $x^k$ are given by the commutators with the Hamiltonian
 \begin{equation}\label{pxdot}
 \dot{p}_j=\frac{q}{mc}g^{kl}{\sf H}_{jk} p_l\,;\qquad
 \dot{x}{\vphantom x}^k=\frac{1}{m}g^{kl}p_l\,.
 \end{equation}
 This system of equations leads to the integrals of motion which can
be determined substituting $\dot{x}$ to $\dot p$ \cite{JL}
 $$
 \frac{\rm d}{{\rm d} t}[p_j-\frac{q}{c}{\sf H}_{jk}x^k]=0\,.
 $$
 Taking into account the definitions (\ref{mag}) and (\ref{pcan}) one
obtains
 \begin{equation}\label{ptcan}
 p_j-\frac{q}{c} {\sf H}_{jk}x^k=
-{\rm i}\hbar\partial_j-\frac{q}{c}{\sf A}^T_{kj}x^k=p_j^T\,.
 \end{equation}
 Due to their definitions, operators $p^T_j$
commute with the Hamiltonian but they do not commute with each other
 \begin{equation}\label{tcomm}
 [p^T_j,p^T_k] = {\rm i}\hbar\frac{q}{c}{\sf H}^T_{jk}
= -{\rm i}\hbar\frac{q}{c}{\sf H}_{jk}\,.
 \end{equation}
 Note that for an antisymmetric gauge, ${\sf A}={\sf H}/2$,
one obtains
 $$
 p^T_j=-{\rm i}\hbar\partial_j+\frac{q}{c}{\sf A}_{kj}x^k=
 -{\rm i}\hbar\partial_j+\frac{q}{c}A_j\,,
 $$
 what agrees with the definitions of magnetic translations used by
Brown and Zak \cite{EB,Zab}. 
 The above presented form
of $p_j^T$ allows us to consider any linear gauge ${\bf A}$ and to
define the magnetic translations as unitary
operators \cite{EB,BH,F7}
 \begin{equation}
 T({\bf x})=\exp[-\frac{\rm i}{\hbar} x^j p^T_j]\,.
 \end{equation}

\section{Cyclotronic orbits}\label{CycOrb}

 The definition (\ref{ptcan}) of $p_k^T$ can be written as a set of
equations for variables $x^l$
 \begin{equation}\label{seteq}
 {\sf H}_{jk}x^k=\frac{c}{q}(p_j-p_j^T)\,.
 \end{equation}
 Since {\sf H} is an antisymmetric matrix then 
its eigenvalues are equal to 0 or are arranged in pairs
of imaginary numbers $\pm\chi_l {\rm i}$, $l=1,2,\dots,N\le n/2$.
Therefore, there exists such an orthogonal basis $\{\varepsilon_\alpha\}$, 
$\alpha=1,2,\dots,n$, in which the magnetic tensor consists of $N$ 
two-dimensional antisymmetric blocks 
 \begin{equation}\label{block}
 \left(\begin{array}{rr}0&\chi_l\\-\chi_l&0\end{array}\right)\,.
 \end{equation}
 Note that this basis can be obtained by a transformation {\sf B} which is
orthogonal with respect to a positive definite metric tensor 
$\gamma_{\alpha\beta}$ not with respect to a general metric tensor $g_{jk}$.
It means that on can choose the basis $\{{\bf e}_j$\} to be orthogonal with
respect to both forms, i.e. $({\bf e}_j)^l({\bf e}_k)^mg_{lm}=g_{jk}$ and 
$({\bf e}_j)^l({\bf e}_k)^m\gamma_{lm}=\gamma_{jk}$ but a new basis
$\varepsilon_\alpha={\sf B}_\alpha^j{\bf e}_j$ is orthogonal, in general,
with respect to the tensor $\gamma$ only. For example, in the most interesting
case when $g_{jj}=1$ for $j<n$, $g_{nn}=-1$ and 0 in the other cases with
$\gamma_{\alpha\beta}=\delta_{\alpha\beta}$,
corresponding to an $(n+1)$-dimensional space-time,
the orthogonal transformation {\sf B}
leads to a basis in which spatial and time coordinates are mixed and
some basis vectors $\varepsilon_\alpha$ can be singular with respect to 
$g_{jk}$.

For the sake of clarity the coordinates of tensors $({\sf H}_{jk})$, $(x^k)$ and
$(p^{(T)}_j)$ in the new basis will be denoted by the Greek letters with
Greek indices, i.e.\ $\Theta_{\alpha\beta}$, $\xi^\beta$ and
$\pi^{(T)}_\alpha$, respectively. Using this notation one can 
 obtain (to simplify notation the factor $c/q$ is included in the defintion
 of $\pi^{(T)}$)
 $$
 \xi^{\beta}={\sf B}^{\beta}_kx^k\,;\qquad
 \pi^{(T)}_{\alpha}=\frac{c}{q}{\sf B}^j_{\alpha}p^{(T)}_j\,;\qquad
{\Theta}_{\alpha\beta}={\sf B}_{\alpha}^jH_{jk}{\sf B}_{\beta}^k\,.
 $$
The $l$th block of the matrix $\Theta_{\alpha\beta}$,
with $l=1,2,\dots,N\le n/2$, 
is given by (\ref{block}), so
the set of equations (\ref{seteq}) can be rewritten as
 \begin{equation}\label{seteqt}
 {\Theta}_{\alpha\beta}\xi^{\beta}=\pi_{\alpha}-\pi_{\alpha}^T\,.
 \end{equation}
 A form of the matrix $\Theta$ leads a decomposition
into two subsets of equations: for odd coordinates $\xi^{2l-1}$  with even
coordinates $\pi_{2l}$ and {\it vice versa}
 \begin{eqnarray}
 \chi_l \xi^{2l-1}&=&-(\pi_{2l}-\pi^T_{2l})\label{eqodd}\,,\\
 \chi_l \xi^{2l}&=&\phantom{-}(\pi_{2l-1}-\pi^T_{2l-1})\label{eqeven}\,.
 \end{eqnarray}
 Therefore, the case $\chi_l\neq0$ corresponds to a movement of a
particle in the plane determined by the vectors $\xi^{2l-1}$ and
$\xi^{2l}$.  The center of the orbit is given by a pair
 \begin{equation}\label{cenorb}
 (\pi^T_{2l},-\pi^T_{2l-1})/\chi_l
 \end{equation}
 and the relative coordinates are
 \begin{equation}\label{relcoo}
 (-\pi_{2l},\pi_{2l-1})/\chi_l\,.
 \end{equation}
 The number of independent cyclotronic movements is equal to $N\le
n/2$, i.e.\ the number of non-zero values of $\chi_l$, and each of
them is described by the Hamiltonian ${\cal
H}_l=(\pi_{2l-1}^2+\pi_{2l}^2)/2m$, $l=1,2,\dots,N$. Of course, it
enables us to introduce the well-known and useful notation
$\pi_l^{\pm}= \pi_{2l-1}\pm{\rm i}\pi_{2l}$. Movements in the other
directions are completely free, i.e.\ they are described by the
Hamiltonian ${\cal H}_{\rm free}=\sum_{\gamma=2N+1}^n
(\dot{\xi}^\gamma\vphantom{x^j})^2/2m$. Using the determined matrix
{\sf B} one can express these Hamiltonians and the center coordinates
in the variables $x^k$ and $p_j$. Due to commutation relations
(\ref{comm}) and (\ref{tcomm}) each cyclotronic movement is
quantized, whereas the other part of the energy spectrum is
continuous.

\section{Conclusions}

It has been indicated that the movement of a charged particle 
in a magnetic field can always be 
decomposed into $N$ cyclotronic motions and $n-2N$ 
free ones. This indicates the differences in the energy
spectrum for $n=2$ and $n=3$ 
 and suggests that for $n=4$ the whole energy
spectrum can be discrete. However, considering the four-dimensional
space-time it means that an electric field has to be included and the metric
tensor of this space is not positively definite. If a
magnetic field is applied only, then there is always such a basis 
in which the tensor $\Theta$ has only one block (\ref{block});
of course this block determines a surface perpendicular to the 
magnetic field ${\bf H}$.

\section*{Acknowledgments}
 This work is partialy supported by the State Committee for Scientific 
Research (KBN) within the project No 263/P03/99/16.

\end{document}